\documentclass[aps,prl,twocolumn,nofootinbib,preprintnumbers]{revtex4}

\usepackage[normalem]{ulem}
\usepackage{amsmath}
\usepackage{enumerate}
\usepackage{amsfonts}
\usepackage{yfonts}

\usepackage{subfigure}
\usepackage{psfrag}

\usepackage{epsfig}
\usepackage[latin1]{inputenc}
\usepackage{float}
\usepackage{graphicx}
\usepackage{cancel}
\usepackage{mathrsfs}
\usepackage{amssymb}
\usepackage{amsfonts}
\usepackage{amsmath}
\usepackage{slashed}
\usepackage{mathtools}

\usepackage{color}

\newcommand{\be}{\begin{equation}}
\newcommand{\ee}{\end{equation}}
\newcommand{\bea}{\begin{eqnarray}}
\newcommand{\eea}{\end{eqnarray}}
\newcommand{\nn}{\nonumber}

\newcommand{\la}{\langle}
\newcommand{\ra}{\rangle}

\newcommand{\lb}{\left(}
\newcommand{\rb}{\right)}

\newcommand{\ep}{\epsilon}
\newcommand{\mR}{{\mathcal R}}

\newcommand{\mO}{{\mathcal O}}
\newcommand{\p}{\partial}

\begin{document}

\title{Modular Hamiltonian of Excited States in Conformal Field Theory}

\author{Nima Lashkari}
\affiliation{Center for Theoretical Physics,
Massachusetts
Institute of Technology,
Cambridge, MA 02139 }

\begin{abstract}
We present a novel replica trick that computes the relative entropy of two arbitrary states in conformal field theory.
Our replica trick is based on the analytic continuation of partition functions that break the $Z_n$ replica symmetry. It provides a method for computing arbitrary matrix elements of the modular Hamiltonian corresponding to excited states in terms of correlation functions. We show that the quantum Fisher information in vacuum can be expressed in terms of two-point functions on the replica geometry.
We perform sample calculations in two-dimensional conformal field theories. 
%
\end{abstract}
 
\maketitle
In recent years entanglement theory has found numerous applications in the study of quantum phases of matter, relativistic  field theories and gravity. Most of these applications focus on an entanglement measure in bi-partite pure states known as the entanglement entropy. Unfortunately, in relativistic field theories entanglement entropy suffers from ultraviolet divergences. In gauge theories the definition of entanglement entropy is ambiguous \cite{Casini:2013rba}. In this letter, we present a method to compute in field theory another measure called relative entropy that is provably ultraviolet finite, universal and free of gauge ambiguities \cite{Araki:1976zv,Casini:2013rba}. 

Relative entropy is a measure of distinguishability between two states and has nice monotonicity and positivity properties. It appears naturally in the definition of entanglement measures for mixed states such as mutual information and the relative entropy of entanglement \cite{Vedral:2002zz}.  Recently, thinking in terms of relative entropy in quantum field theories coupled to gravity has led to new developments such as a proof of quantum Bousso bound \cite{Bousso:2014sda}, and the identification of new gravitational positive energy theorems \cite{Lashkari:2016idm}. 

The relative entropy of the density matrix $\phi$ with respect to $\psi$ is defined to be 
\bea
S(\phi\|\psi)=tr(\phi\log\phi)-tr(\phi\log\psi). 
\eea
Note that relative entropy is ill-defined when $\psi$ is pure.
The relative entropy of two states can be thought of as the expectation value of the difference of the modular Hamiltonians of the two states
\bea
S(\phi\|\psi)&=&\la\phi|H(\psi)-H(\phi)|\phi\ra\nn\\
&=&tr((\phi-\psi)H(\psi))-\Delta S.
\eea
Here the positive Hermitian operator $H(\psi)=-\log\psi$ is the modular Hamiltonian of $\psi$, and $\Delta S$ is the difference of the entanglement entropies of $\phi$ and $\psi$. If we formally define the {\it generalized free energy} function $F_\psi(\phi)=tr(\phi H(\psi))-S(\phi)$, then the relative entropy is the free energy difference between the two states
\bea
S(\phi\|\psi)=F_\psi(\phi)-F_\psi(\psi).
\eea
The function $F_\psi$ has all the properties one expects from free energy in a thermodynamic theory where $\psi=e^{-H(\psi)}$ plays the role of the equilibrium state \cite{brandao:resource,Lashkari:2016idm}. Note that $F_\psi$ achieves its minimum on the equilibrium state $\psi$.\footnote{This is a consequence of positivity of relative entropy: $F_\psi(\phi)\geq F_\psi(\psi)$.}

In this letter, we construct a class of field theory partition functions that are proportional to $tr(\phi \psi^{n-1})$. Their analytic continuation provides the relative entropy and the modular Hamiltonian of density matrices in excited states $|\phi\ra$ and $|\psi\ra$ reduced to the subsystem. While the formalism presented here applies to all quantum field theories we specialize to conformal field theories to have access to more computational tools.

\begin{figure}
\centering
\includegraphics[width=0.4\textwidth]{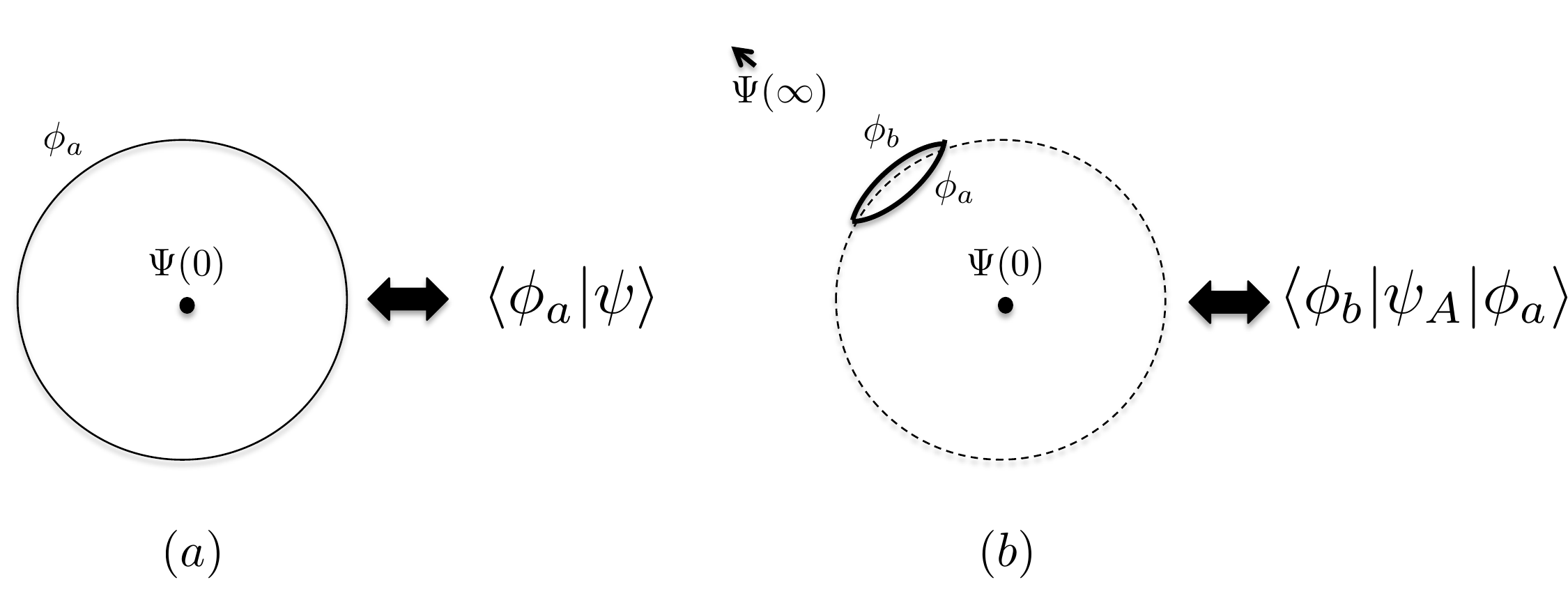}\\
\caption{\small{($a$) Operator-state correspondence in radial quantization of conformal field theories. ($b$) Reduced density matrix corresponds to a path-integral with two operator insertions and a cut on the subsystem.}}
\label{fig1}
\end{figure}

 According to the operator-state correspondence in conformal field theory (CFT) there is a one-to-one map between wave-functionals and operators in the Hilbert space. In radial quantization, the wave-functional of an excited state $|\psi\ra$ is found by performing a Euclidean path-integral with the corresponding operator $\Psi$ inserted. Restricting to subsystem $A$ the state is described by a density matrix $\psi_A$; see figure \ref{fig1}. To simplify notation we suppress the subsystem index $A$, and use $\psi$ to refer to the reduced state. 
 
 In principle, one can compute the logarithm of density matrix directly from the path-intergral by taking the logarithm of a path-ordered operator using the so-called Magnus expansion, however in practice this is too hard. Here, we propose an alternative method to compute matrix elements of the modular Hamiltonian of excited states from the analytic continuation of correlation functions. Our method is a generalization of the replica trick in \cite{Holzhey:1994we,Calabrese:2004eu} to the case where one breaks the $Z_n$-symmetry among replicas. This enables us to compute matrix elements of the modular operator for all states. This is in contrast with the old replica method which was restricted to states that have local modular Hamiltonians, the only known example of which is vacuum reduced to a half-space or spherical subsystems \cite{Unruh:1976db,Casini:2011kv}.\footnote{In two dimensional CFTs, a finite temperature state on a line also has reduced states with local modular Hamiltonian \cite{Wong:2013gua}.} 

\begin{figure}
\centering
\includegraphics[width=0.5\textwidth]{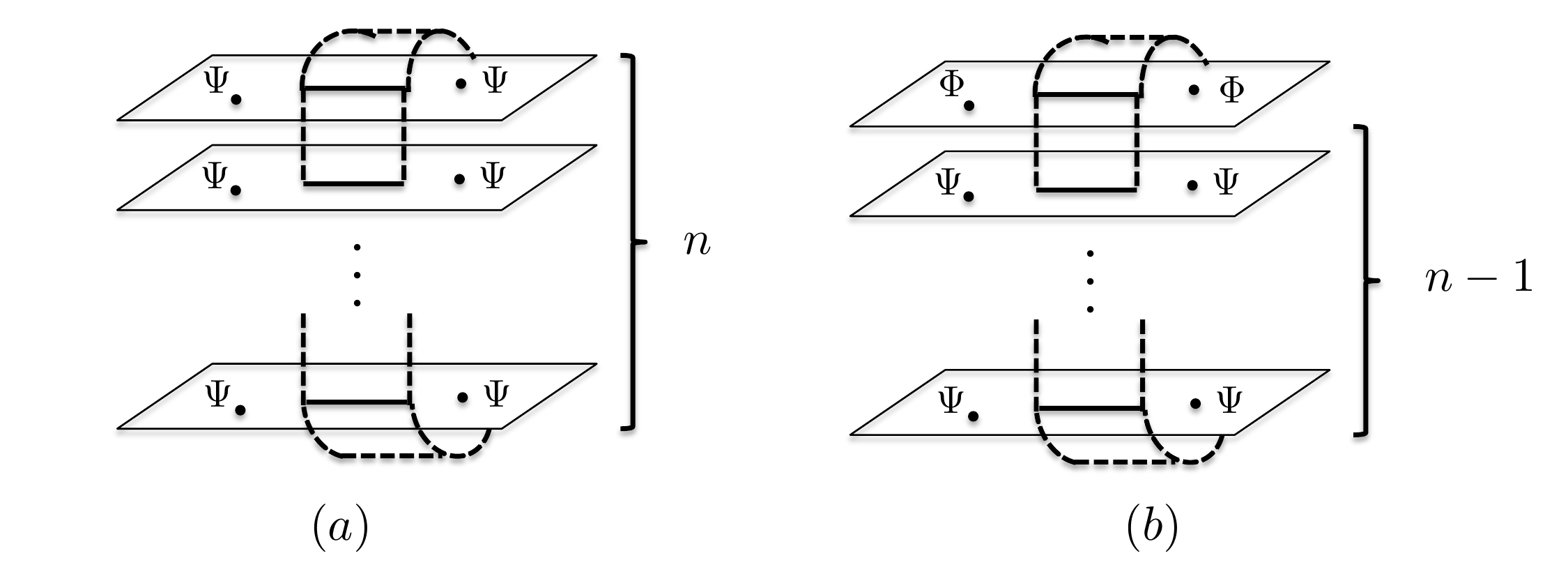}\\
\caption{\small ($a$) Entanglement entropy replica trick: the Euclidean path-integral on the $n$-sheeted manifold corresponding to the partition function $tr(\psi^n)$. ($b$) The $Z_n$-breaking partition $tr(\phi\psi^{n-1})$ that appears in our relative entropy replica trick.} \label{fig2}
\end{figure}

\section{Relative entropy and Modular Hamiltonian} 

Consider the Hermitian operator $\{\phi,\psi^{n-1}\}=\frac{1}{2}(\phi\psi^{n-1}+\psi^{n-1}\phi)$ built out of reduced density matrices $\phi$ and $\psi$ corresponding to global states $|\phi\ra$ and $|\psi\ra$, respectively. Its trace in conformal field theory corresponds to the $n$-sheeted partition function $tr(\phi\psi^{n-1})$. The idea is to take advantage of the analytic properties of correlators by using the operator identity 
\bea
tr(\phi\log\psi)=\p_n(\phi\psi^{n-1})\big|_{n\to 1}.
\eea
Our partition functions of interest, $tr(\phi\psi^{n-1})$, break the $Z_n$ replica symmetry present in both the Renyi entropy and Renyi relative entropy replica tricks \cite{Calabrese:2004eu,Lashkari:2014yva}. In contrast with the symmetric case, our partition functions are not monotonic in index $n$, and have no known operational interpretations. Nonetheless, under the assumption of analyticity, they provide a computational tool for finding relative entropies and the diagonal elements of the modular Hamiltonian of excited states
\bea\label{rel}
&&{\mathcal S}(\phi\|\psi)=\p_n\log\left[ \frac{tr(\phi^n)tr(\psi)^{n-1}}{tr(\phi\psi^{n-1})tr(\phi)^{n-1}}\right]_{n\to1}\nn\\
&&\la \phi| H(\psi)-H(\sigma_0) |\phi\ra=\log\left[\frac{\la \phi|\sigma_0^{n-1}|\phi\ra tr(\psi)^{n-1}}{\la\phi|\psi^{n-1}|\phi\ra tr(\sigma_0)^{n-1}} \right]_{n\to1},\nn\\
\eea
where $\sigma_0$ is the reduced density matrix in vacuum. We subtract the vacuum modular Hamiltonian so that we have ultra-violet finite quantities at any $n$. The off-diagonal elements of the modular Hamiltonian are obtained from its diagonal element in superposition states; see supplementary material. 

Each of the terms inside the logarithm above can be expressed as a Euclidean path-integral with operator insertions on a replicated or the original geometry \cite{Alcaraz:2011tn}. For instance, consider the terms $tr(\rho\psi^{n-1})$. Sewing $n$ copies of the density matrix cyclically along the boundary of their subsystems we obtain $n$-sheeted replica manifold $\mR_n$ and $2n$ operator insertions (figure \ref{fig2})
\bea\label{operationinsert}
&&tr(\phi\psi^{n-1})=Z({\mR_n}) \la\Phi(z'_n)\Phi(z_n)\mO^{(n-1)}_\Psi\ra_{\mR_n}\nn\\
&&\mO^{(m)}_\Phi=\prod_{i=1}^{m}\Phi(z'_i)\Phi(z_i)
\eea
where $z_i$ and $z'_i$ are points $z$ and $z'$ on the $i^{th}$ sheet of $\mR_n$. It is important to note that plugging (\ref{operationinsert}) in (\ref{rel}) all partition function terms cancel and we are left only with correlation functions at any $n$ which are free of ultraviolet divergences.


Written explicitly in terms of correlation functions we find the main results of this section:
\bea\label{main}
&&S(\phi\|\psi)=\nn\\
&&\p_n\log\left[ \frac{\la \mO^{(n)}_{\Phi} \ra_{\mR_n}\la\Psi(z'_1)\Psi(z_1)\ra_{\mR_1}^{n-1}}{\la\Phi(z'_n)\Phi(z_n)\mO^{(n-1)}_\Psi\ra_{\mR_n}\la \Phi(z'_1)\Phi(z_1)\ra_{\mR_1}^{n-1}}\right]_{n\to 1}\nn\\
&&\la \phi| H(\psi)-H(\sigma_0) |\phi\ra=\nn\\
&&\p_n\log\left[ \frac{\la \Phi(z'_n)\Phi(z_n)\ra_{\mR_n}\la\Psi(z'_1)\Psi(z_1)\ra_{\mR_1}^{n-1}}{\la\Phi(z'_n)\Phi(z_n)\mO^{(n-1)}_\Psi\ra_{\mR_n}}\right]_{n\to 1}\nn\\
&&\la\chi|H(\psi)-H(\sigma_0)|\phi\ra=\p_n\left[\log\left[\frac{X_{-1}}{X_{+1}}\right]+i\log\left[\frac{X_{-i}}{X_{+i}}\right]\right]_{n\to 1}\nn
\eea
where 
\bea
&X_c&= E_{\Phi\Phi}(\mO_\Psi^{(n-1)})+|c|^2\:E_{\chi\chi}(\mO_\Psi^{(n-1)}) \nn\\
&&+c\: E_{\Phi\chi}(\mO_\Psi^{(n-1)})+h.c.\nn\\
&E_{\Phi\chi}&(\mO)=\frac{\la \Phi(z'_n)\chi(z_n) \mO\ra_{\mR_n}}{\sqrt{\la \Phi(z'_1)\Phi(z_1)\ra_{\mR_1}\la \chi(z'_1)\chi(z_1)\ra_{\mR_1}}}
\eea
Here, we have assumed that $\la\psi|\phi\ra=0$.



\section{Quantum Fisher information}
Our replica trick connects the modular Hamiltonian of excited states to analytic continuation of $2n$-point correlation functions. Apart from integrable models and large central charge theories obtaining analytic expressions for $2n$-point functions is an intractable problem. However, as we show in this section a great simplification occurs once we focus on near-vacuum states. 

Let us first consider a one-parameter family of states $\frac{(|\phi\ra+\ep |X\ra)}{\sqrt{1+\ep^2}}$ perturbed around $|\phi\ra$ in perpendicular direction $|X\ra$. The reduced density matrix on subsystem $A$ expanded in $\ep$ has the form 
\bea
&&\phi+\ep \rho_X^{(1)}+\ep^2 \rho_X^{(2)}+O(x^3)\nn
\eea
Relative entropy is a smooth non-degenerate function of two states. Hence, the relative entropy of two nearby states expanded in $\ep$ vanishes to the first order. 
The coefficient of the second order term, $F_\phi(X,Y)$, is called the quantum Fisher information at point $\phi$ in the space of density matrices:
\bea
S(\phi+\ep \rho_X\|\phi)=\ep^2 F_\phi(X,X)+O(\ep^3).\nn
\eea
This function defines a metric on the space of perturbations to state $\phi$
\bea
2F_\phi(X,Y)=F_\phi(X+Y,X+Y)-F_\phi(X,X)-F_\phi(Y,Y).\nn
\eea
Quantum Fisher information is a local measure of distinguishability, and is intimately connected with uncertainty relations \cite{Petz}. 
Consider the relative entropy of two nearby states. Our replica trick in (\ref{rel}) implies 
\bea\label{FishgerGen}
&&F_\phi(X,X)=\nn\\
&&\p_n\left[n\sum_{m=0}^{n-2}\frac{\la \{X,\Phi\}\Phi^{2m} \{X,\Phi\}\Phi^{2(n-m-2)}\ra_{\mR_n}\la \Phi\Phi\ra_{\mR_1}}{\la \Phi^{2n}\ra_{\mR_n}\la X X\ra_{\mR_1}}\right]_{n\to 1}\nn\\
\eea
 where $X$ and $\Phi$ denotes the operators that create the perturbations corresponding to $|X\ra$ and $|\phi\ra$, respectively. The location of operator insertions are the same as (\ref{operationinsert}). 

\begin{figure}
\centering
\includegraphics[width=0.5\textwidth]{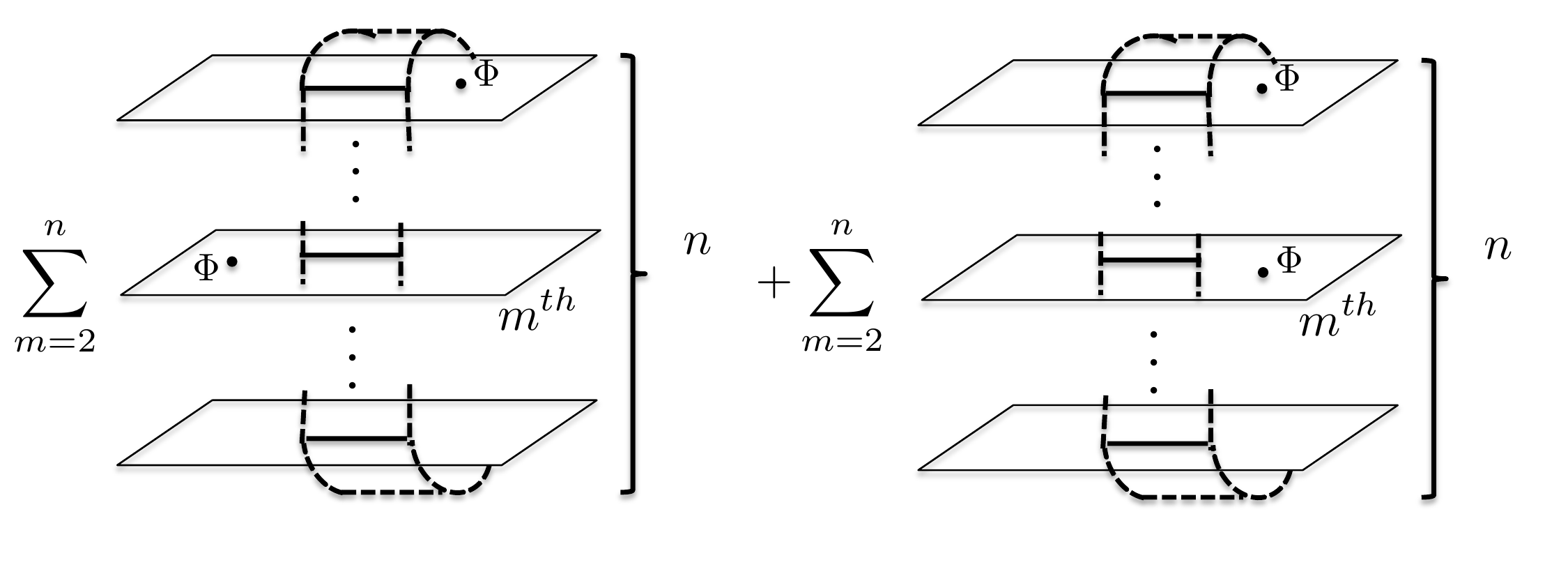}\\
\caption{\small The type of two-point functions on the replica manifold whose analytic continuation determine quantum Fisher information in vacuum.} \label{fig3}
\end{figure}

For near vacuum states, we replace $\Phi$ in (\ref{FishgerGen}) with identity. The quantum Fisher information takes the form of an analytic continuation of two-point functions on the replica geometry; see figure \ref{fig3}:
 \bea\label{Fisheruniv}
 &&F_\sigma(X,X)=\p_n\left[\sum_{z^\pm=z,z'}K_X(z^+,z^-)\right]_{n\to 1}\nn\\
 &&K_X(z^+,z^-)=n\sum_{m=1}^{n-1}\frac{ \la X(z_1^+)X(z_{m+1}^-)\ra_{\mR_n}}{\la X(z_0)X(z'_0)\ra_{\mR_1}}.
 \eea

 This implies that in arbitrary dimensions the vacuum Fisher information of any primary excitation reduced to a ball or radius $R$ is universal in the sense that it depends only on energy and subsystem size. 
In the remaining of this letter, we provide examples of relative entropies, modular Hamiltonians and quantum Fisher information in two-dimensional CFTs computed using the method above.

\section{Examples in two dimensions}

\paragraph*{Relative entropy of excited states:}
Consider a free massless boson CFT in two dimensions on a circle of radius $R$ and a subsystem at $A=(-l/2,l/2)$. We are interested in the excited states obtained by the action of chiral vertex operators on vacuum at past infinity: $|\alpha\ra=V_\alpha|\Omega\ra=e^{i\alpha  \phi}|\Omega\ra$, where $\phi$ is the boson field. The dimension of this operator is $(h,\bar{h})=(\alpha^2/2,0)$. Here $x=l/R$ is the dimensionless parameter. 
In supplementary material it is shown that in two dimensions one can equally use correlators on cylinder, full complex plane or a strip in our formulae in (\ref{main}) for relative entropy and modular Hamiltonian. The conformal factors found from the change of coordinates vanish in the limit of $n\to 1$.
In a free theory with a non-degenerate ground state all correlation functions are determined by Wick's theorem \cite{yellowbook}: 
%
\bea
&&\la V_{-\alpha}\mO^{(n-1)}_\beta V_\alpha\ra_S
=(2\sin(\pi x/n))^{\beta^2(1-n)-\alpha^2}g_n^{-(n-2)\beta^2-2\alpha\beta}\nn
\eea
where $S$ refers to correlators on a strip of width $2\pi$, and $g_n=\frac{\sin(\pi x)}{n\sin(\pi x/n)}$.
For holomorphic excitations $[V_\alpha]^\dagger\sim V_{-\alpha}$. Therefore,
\bea\label{relfree}
&&S(\alpha\|\beta)=\p_n\log\lb \frac{\la\mO^{(n)}_\alpha \ra_S\la V_{-\beta}V_{\beta}\ra^{n-1}_S}{\la V_{-\alpha}V_\alpha\mO^{(n-1)}_\beta\ra_S\la V_{-\alpha}V_\alpha\ra^{n-1}_S}\rb\nn\\
&&=(\alpha-\beta)^2(1-\pi x \cot(\pi x)).
\eea
The analytic continuation used above is justified in supplementary material. When $\beta=0$ this matches the result previously found using a $Z_n$-symmetric replica trick in \cite{Lashkari:2014yva}: $S(\alpha\|0)=\alpha^2(1-\pi x\cot(\pi x))$. Interestingly, the answer in (\ref{rel}) is symmetric in its arguments, $S(\alpha\|\beta)=S(\beta\|\alpha)$. These excited states further have the property that $S(\alpha)=S(\beta)=S(\sigma_0)$, where $\sigma_0$ is the vacuum density matrix. Hence, we find $tr(\rho_\alpha H_\beta)=tr(\rho_\beta H_\alpha)$ for all $\alpha$ and $\beta$. 


\paragraph*{Modular Hamiltonian of excited states:}
In the free $c=1$ CFT Wick contractions imply that a correlator is zero unless $\sum_i\alpha_i=0$. For all $\alpha\neq \gamma$ we have $\la V_{-\alpha}V_\gamma\mO^{(n-1)}_\beta\ra_S=0$. As a result, $X_c$ in (\ref{main}) is independent of $c$, and we find that the modular operator $H_\beta$  has no off-diagonal terms in the $|\alpha\ra$ basis. 

 The diagonal elements are
 \bea
 \la \alpha| H(\beta)-H(\sigma_0)|\alpha\ra
 =\beta(\beta-2\alpha)(1-\pi x\cot(\pi x)).\nn
 \eea
 Note that in the limit $\alpha=\beta$ this reproduces $-\beta^2(1-\pi \cot(\pi x))=-S(\beta|\sigma_0)$ as it should. In the limit $\beta=0$ the difference of modular Hamiltonians is the zero operator and hence the answer should vanish as it does.

\paragraph*{Quantum Fisher metric around the vacuum:} Consider an arbitrary two-dimensional conformal field theory on a circle. The vacuum Fisher information is given by equation (\ref{Fisheruniv}). 
After some algebra we find
\bea
&&F_\sigma(X,X)=\p_n\left[2s_n(0)+s_n(x)+s_n(-x)\right]_{n\to 1}\nn\\
&&s_n(x)=\lb\frac{\sin^2(\pi x)}{n^2}\rb^{h+\bar{h}}n \sum_{m=1}^{n-1}\sin\lb\pi (m+x)/n \rb^{-2(h+\bar{h})}\nn
\eea
For simplicity we expand in small $x$ to find:
\bea
F_\sigma&&\simeq\p_n\left[4\lb\frac{\pi x}{n}\rb^{2(h+\bar{h})}n \sum_{m=1}^{n-1}\sin(\pi m/n)^{-2(h+\bar{h})}\right]_{n\to 1}\nn\\
&&=\frac{2 (\pi x)^{2(h+\bar{h})}\sqrt{\pi}\Gamma(h+\bar{h}+1)}{\Gamma(h+\bar{h}+\frac{3}{2})},
\eea
where we have used the analytic continuation found in \cite{Calabrese:2010he}.

\begin{figure}
\centering
\includegraphics[width=0.5\textwidth]{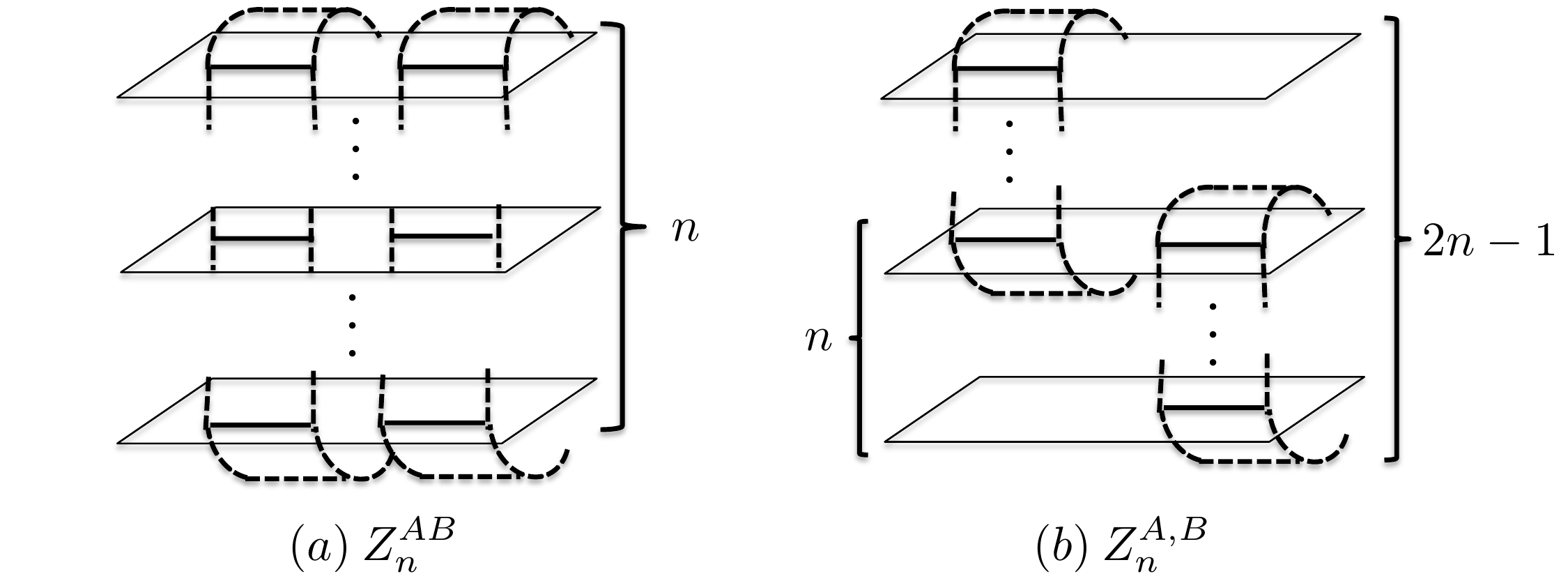}\\
\caption{\small ($a$) The $n$-sheeted manifold corresponding to the partition function $Z_n^{AB}$. ($b$) The $Z_n$-breaking partition $Z_n^{A,B}=tr(\sigma_{AB} (\sigma_A\otimes \sigma_B)^{\otimes n-1})$.} \label{fig4}
\end{figure}

\begin{figure}
\centering
\includegraphics[width=0.5\textwidth]{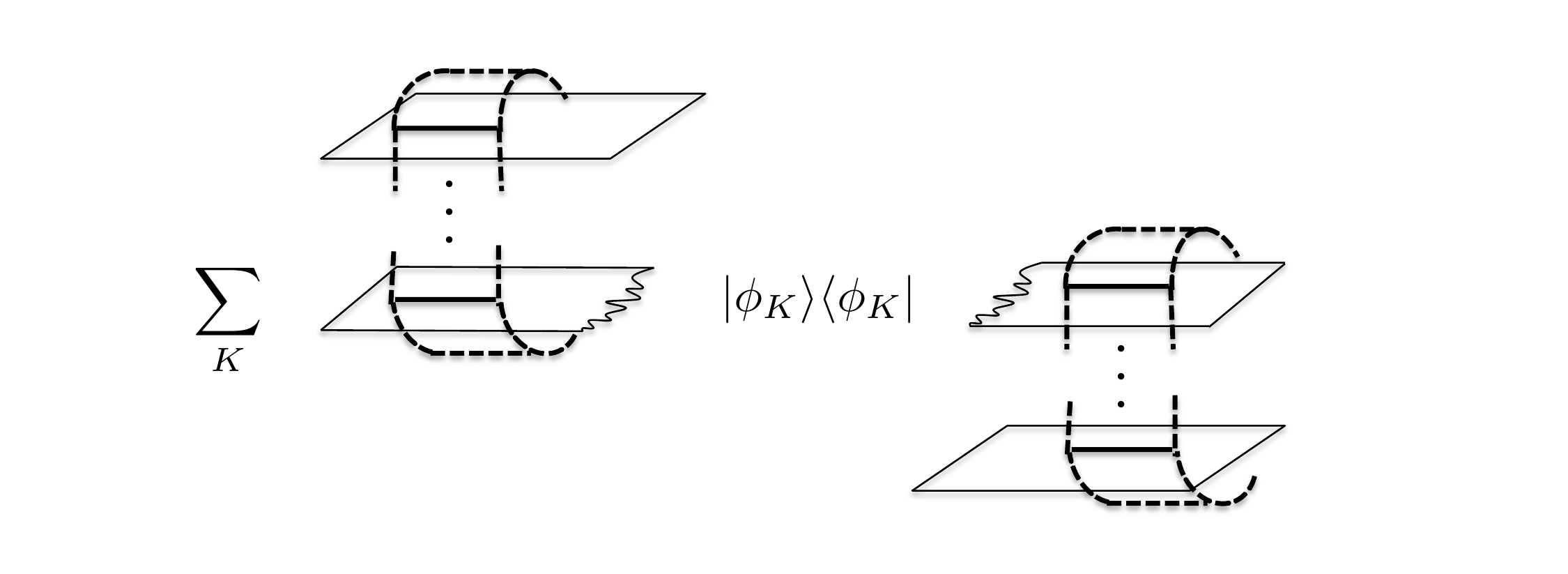}\\
\caption{\small Inserting the resolution of identity in $Z_n^{AB}$ we observe that at each $K$ we multiply sphere one-point functions that are zero unless $\Phi_K$ is identity.} \label{fig5}
\end{figure}

\paragraph*{Multiple intervals:} The replica trick developed here can be applied to subsystems with multiple intervals. As an example we focus on mutual information in vacuum  
\bea\label{mutual}
S(\sigma_{AB}\|\sigma_A\otimes \sigma_B)=I(A:B),
\eea
 where $A$ and $B$ are non-overlapping intervals.  According to (\ref{rel}) the relative entropy is the analytic continuation of the vacuum partition functions on manifolds $Z_n^{AB}$ and $Z_n^{A,B}$ illustrated in figure \ref{fig4}
 \bea
 I(A:B)=\lim_{n\to 1}\frac{1}{n-1}\lb\log Z_n^{AB}-\log Z_n^{A,B}\rb.
 \eea
 The first partition function $Z_n^{AB}$ corresponds to Renyi entropies of $\sigma_{AB}$. Therefore, from (\ref{mutual}) all we need to check is 
 \bea\label{sumentr}
 \p_nZ_n^{A,B}\big|_{n\to 1}=S(A)+S(B).
 \eea
 Riemann-Hurwitz formula tells us that $Z_n^{AB}$ has genus $(n-1)$ and $Z_n^{A,B}$ is simply the Riemann sphere. Following \cite{Lunin:2001ew} we compute the path-integral over these manifolds using twist operators in an orbifold theory with replica copies of the fields. In particular, up to normalization $Z_n^{A,B}$ is the correlation function 
 \bea
\la \sigma_{(1\cdots n)}(u_A)\sigma_{(n\cdots 1)}(v_A)\sigma_{(n\cdots 2n-1)}(u_B)\sigma_{(2n-1\cdots n)}(v_B)\ra\nn
 \eea
 in a $(2n-1)$ replica theory.  Here $u_A$ and $v_A$ are the endpoint of interval $A$, and going around the twist operator $\sigma_{(1\cdots m)}$ the replica fields transform as $(X^1,X^2\cdots X^m,X^{m+1}\cdots X^{2n-1})\to (X^2,\cdots X^m,X^1,X^{m+1}\cdots X^{2n-1})$. Inserting a resolution of identity on $n^{th}$ sheet splits the correlator into a sum over the product of sphere one-point functions. (see figure \ref{fig5}). Sphere one-point function are zero unless $\Phi_k$ is the identity operator. In other words,
 \bea
 S_n^{A,B}=S_n^{A}+S_n^{B}
 \eea 
 which is the sum of Renyi entropies of intervals $A$ and $B$; and hence (\ref{sumentr}) follows. 

\section{Discussion}
In this letter, we have developed a replica trick that takes advantage of breaking the replica symmetry to access the modular Hamiltonian of excited states. In the absence of the $Z_n$ replica symmetry {\it Renyi}'s are not monotonic in $n$, hence our method cannot be used to obtain lower or upper bounds on relative entropy. The applicability of this method crucially relies on our ability to analytically continue correlation functions in $n$. According to the Carlson theorem \cite{Carlson}, in order to find the unique analytic continuation of Renyi's at integer $n$ one needs to further fix the behavior at $n\to \pm i \infty$. We postpone a careful study of this asymptotic choice and its physical implications to future work. 

The correlation functions needed to compute the modular operator of an excited state are $2n$-point functions. There are not many examples of CFTs for which we have access to high-point correlators. One class of such CFTs are free theories which we briefly discussed. Another class are CFTs with large central charge, where one can reduce the calculation of $n$-point functions of heavy operators to a classical monodromy problem for  differential equations that correlation functions satisfy \cite{Belavin:1984vu}.

In holographic theories, the vacuum Fisher information in spherical subsystems was recently shown to be dual to canonical energy in gravity \cite{Lashkari:2015hha}. This confirms the universal feature suggested by equation (\ref{Fisheruniv}). It would be interesting to understand the connection between the CFT calculation of this quantity and canonical energy in the bulk. 

\paragraph*{Acknowledgements:}
We are greatly indebted to Sean Hartnoll whose observation of the importance of partition functions that break replica-symmetry initiated this project. 
We thank Salman Beigi, John Cardy, Thomas Hartman and Srivatsan Rajagopal for discussions and illuminating comments. This work is supported in part by funds provided by MIT-Skoltech Initiative.



%
\appendix
\section{Superposition states}\label{superpose}
In the radial quantization, the wave-function of an excited state is a path-integral over the disk with unit radius and an operator inserted at the center of the disk: 
\bea
\la \phi_a|\phi\ra=\int_{\phi_a}{\mathcal D}\phi(\omega) e^{-S[\phi]} \Psi(\omega=0), 
\eea
where the boundary conditions are $\phi(\omega=1)=\phi^a$.
It is immediate that a superposition state of $|\phi\ra$ and $|\chi\ra$ is given by 
\bea\label{superpose2}
c_\psi \la\phi_a|\psi\ra+c_\chi\la\phi_a|\chi\ra=\int_{\phi_a} {\mathcal D}\phi e^{-S(\phi)} (c_\psi\Psi(0)+c_\chi\chi(0)).\nn
\eea
The complex conjugate state $\la \phi|$ is created by the action of $\left[\phi(z,\bar{z})\right]^\dagger=\bar{z}^{-2h}z^{-2\bar{h}}\phi(1/\bar{z},1/z)$. 
A simple way to keep track of normalization for  superposition states is to normalize term by term
\bea
\la\xi_c|A|\xi_c\ra=\frac{\la \Phi A\Phi\ra}{\la\Phi\Phi\ra}+\frac{\la \chi A\chi\ra}{\la\chi\chi\ra}+\frac{c\la \Phi A\chi\ra}{\sqrt{\la\Phi\Phi\ra\la\chi\chi\ra}}+h.c.\nn
\eea
We are interested in off-diagonal elements of the modular operator which are obtained from its diagonal element in superposition states 
\bea\label{offdia}
&&\la \chi|A|\phi\ra=\frac{1}{2}\lb A_{(1)}-A_{(-1)}+i A_{(+i)}-i A_{(-i)}\rb,
\eea
where $\la \xi_c|A|\xi_c\ra=A_{(c)}$ and $|\xi_c\ra=\frac{1}{\sqrt{1+|c|^2}}(|\phi\ra +c |\chi\ra)$. 
Note that $|\phi\ra$ and $|\chi\ra$ are orthogonal: $\la\chi \Phi \ra=0$.

The expectation value of the modular Hamiltonian in superposition state $\xi_c$ is:
\bea\label{Xc}
&&\la \xi_c|H(\psi)-H(\sigma_0)|\xi_c\ra=\nn\\
&&\log\la \Psi(t'_0)\Psi(t_0)\ra_{\mR_1}-\p_n\log\left[\frac{X_c }{E_{\Phi\Phi}(1)+|c|^2E_{\chi\chi}(1)}\right]_{n\to 1}\nn
\eea
From (\ref{offdia}) we find 
\bea\label{off}
\la\chi|H(\psi)-H_0|\phi\ra=\p_n\left[\log\left[\frac{X_{-1}}{X_{+1}}\right]+i\log\left[\frac{X_{-i}}{X_{+i}}\right]\right]_{n\to 1}.\nn
\eea
where 
\bea
&X_c&= E_{\Phi\Phi}(\mO_\Psi^{(n-1)})+|c|^2\:E_{\chi\chi}(\mO_\Psi^{(n-1)}) \nn\\
&&+c\: E_{\Phi\chi}(\mO_\Psi^{(n-1)})+h.c.\nn\\
&E_{\Phi\chi}&(\mO)=\frac{\la \Phi(z'_n)\chi(z_n) \mO\ra_{\mR_n}}{\sqrt{\la \Phi(z'_1)\Phi(z_1)\ra_{\mR_1}\la \chi(z'_1)\chi(z_1)\ra_{\mR_1}}}.
\eea
 
\section{Conformal map from cylinder to complex plane and strip}\label{toStrip}

There are two choice of coordinates that are more convenient for computations in two-dimensional CFTs: the complex plane coordinates ($\omega$), and the strip coordinates ($t$). Our starting point is the cylinder path-integral with operators inserted at $z=\pm i\infty$ and boundary conditions $\phi_a$ and $\phi_b$ above and below $A$. We first apply the map 
\bea
\omega=\frac{\sin(\pi(z+l/2)/R)}{\sin(\pi (z-l/2)/R)}
\eea 
that sends the cylinder $\mR_1$ to the complex plane. Under this transformation, the subsystem $A$ is mapped to the half-line $(0,\infty)$, and operators insertions are at $e^{\pm i \pi x}$. Then, we apply a second conformal map: $\omega=e^{i t}$. The complex $t$-plane is a strip of width $2\pi$ with boundary conditions imposed on $t=\pm\pi$ lines, and operators inserted at $t=\pm \pi x$. Similar maps
 
\bea\label{trans}
&&\omega=\lb\frac{\sin(\pi(z+l/2)/R)}{\sin(\pi (z-l/2)/R)}\rb^{1/n}\nn\\
&&t=-\frac{i}{n}\log\lb\frac{\sin(\pi (z+l/2)/R)}{\sin(\pi (z-l/2)/R)}\rb
\eea
uniformize the n-sheeted Riemann surface $\mR_n$ to the complex plane and a strip of width $2\pi$, respectively.
For simplicity we focus on primary fields of dimension $(h_\psi,\bar{h}_\psi)$ which transform according to
\bea\label{transf}
\Psi(z,\bar{z} )=\lb\frac{\p z}{\p t} \rb^{-h_\psi}\lb\frac{\p\bar{z}}{\p \bar{t}} \rb^{-\bar{h}_\psi}\Psi(t,\bar{t})
\eea
Then,
\bea\label{transprim}
&&\lb\frac{\p z}{\p \omega}\rb=\frac{-R n}{2\pi\omega} \frac{\sin(\pi x)}{(\cos(i n\log(\omega))-\cos(\pi x))}\nn\\
&&\lb \frac{\p z}{\p t}\rb=\frac{i n R}{2\pi}\frac{\sin(\pi x)}{(\cos(n t)-\cos(\pi x))}.
\eea

Under a change of coordinates (\ref{rel}) transforms as
 \bea
&&S(\phi\|\psi)=\p_n\log\left[ {\mathcal J}_S(\{t_i\},h_\phi,h_\psi){\mathcal J}_S(\{t'_i\},h_\phi,h_\psi)\right]_{n\to 1}\nn\\
&&+\p_n\log\left[ \frac{\la \mO^{(n)}_{\Phi} \ra_{S}\la\Psi(t'_0)\Psi(t_0)\ra_{S}^{n-1}}{\la\Phi(t'_n)\Phi(t_n)\mO^{(n-1)}_\Psi\ra_{S}\la \Phi(t'_0)\Phi(t_0)\ra_{S}^{n-1}}\right]_{n\to 1}\nn
 \eea
 where the subscript $S$ refers to the correlators on a strip, and
  \bea
{\mathcal J}_S(t_i)=&&\lb \frac{\p z}{\p t}\Big|_{n=1}\rb^{(n-1)\Delta}_{t_0}\lb \frac{\p \bar{z}}{\p \bar{t}}\Big|_{n=1}\rb_{\bar{t}_0}^{(n-1)\bar{\Delta}}\nn\\
 &&\times \prod_{i=1}^{n-1}\lb \frac{\p z}{\p t}\rb^{-\Delta}_{t_i}\lb \frac{\p \bar{z}}{\p \bar{t}}\rb^{-\bar{\Delta}}_{\bar{t}_i},
 \eea
 where $\Delta=h_\phi-h_\psi$. 
 On the $t$-plane operator insertions are at $t_i=\pi (2i+x)/n$ and $t'_i=\pi (2i-x)/n$, $t_0=\pi x$, $t'_0=-\pi x$ so (\ref{transprim}) needs to be regulated.  However, the regulators cancel in the expression for ${\mathcal J}_S$:
\bea
{\mathcal J}_S=&&\lb \frac{n R}{2\pi} \frac{\sin(\pi x)}{(\cos(n t)-\cos(\pi x))}\rb^{-2(n-1)(\Delta+\bar{\Delta})}_{t\to \pi x/n}  \nn\\
&&\times\lb\frac{R}{2\pi} \frac{\sin(\pi x)}{(\cos( t)-\cos(\pi x))}\rb^{2(n-1)(\Delta+\bar{\Delta})}_{t\to \pi x}\nn\\
&&=n^{-2(n-1)(\Delta+\bar{\Delta})}
\eea
Hence, in the limit $n\to 1$ there is no contribution to (\ref{rel}) from the change of coordinates to the strip.
 A similar argument shows that the conformal factors do not contribute to the expression for the modular operator either. The same is true for using correlators on the full complex plane in (\ref{transprim}). In other words, we have
\bea\label{main}
&&S(\phi\|\psi)=\p_n\log\left[ \frac{\la\mO^{(n)}_\Phi\ra\la \Psi\Psi\ra^{n-1}}{\la\Phi\Phi\mO^{(n-1)}_\Psi\ra\la \Phi\Phi\ra^{n-1}}\right]_{n\to 1}\nn\\
&&\la \phi|(H(\psi)-H(\sigma_0))|\phi\ra=\p_n\log\left[ \frac{\la \Phi\Phi\ra\la\Psi\Psi\ra^{n-1}}{\la\Phi\Phi\mO^{(n-1)}_\Psi\ra}\right]_{n\to 1}
\eea
where the correlators can be either on the strip or the full complex plane, and we have suppressed the location of operator insertions.

\section{Analytic continuation}\label{analytic}
We are interested in finding the analytic continuation of following finite sums
\bea
\prod_{m=1}^{n-1} f_{n,m}(x)^{n-m},\quad \prod_{m=1}^{n-2} f_{n,m}(x)^{n-m-1},\quad \prod_{m=1}^{n-1} f_{n,m}(x),\nn
\eea
where $f_{n,m}$ is given by
\bea
f(n,m)=\frac{\sin^2(\pi m/n)}{\sin(\pi (m+x)/n)\sin(\pi(m-x)/n)}.
\eea 
The answer is given by the identities below:
\bea
&&\prod_{m=1}^{n-1} f_{n,m}(x)^{n-m}=\lb \frac{n\sin(\pi x/n)}{\sin(\pi x)}\rb^n\nn\\
&&\prod_{m=1}^{n-2} f_{n,m}(x)^{n-m-1}=\lb \frac{n\sin(\pi x/n)}{\sin(\pi x)}\rb^{n-2}\nn\\
&&\prod_{m=1}^{n-1} f_{n,m}(x)=\lb \frac{n\sin(\pi x/n)}{\sin(\pi x)}\rb^2\nn\\
&&\prod_{m=1}^{n-1}\frac{\sin^{(\alpha+\gamma)\beta}(\pi m/n)}{\sin^{\alpha\beta}(\pi (m+x)/n)\sin^{\gamma\beta}(\pi(m-x)/n)}\nn\\
&&=\lb \frac{n\sin(\pi x/n)}{\sin(\pi x)}\rb^{(\alpha+\gamma)\beta}
\eea
To prove these identities, first bring the expression on the right to the left-hand side. Then, note that both the numerator and denominator always have poles of the same degree, and therefore has no poles at any value of $x$. Taking the limit $x\to 0$ we find that our four expressions are equal to one for all $x$, $n$, $\alpha$, $\beta$ and $\gamma$. 

\end{document}